\documentclass[conference]{IEEEtran}
\usepackage[margin=0.75in,top=0.75in,bottom=1in]{geometry}
\IEEEoverridecommandlockouts                              

\usepackage{cite}
\usepackage{amsmath,amssymb,amsfonts}

\usepackage{graphicx}
\usepackage[small]{caption} 
\usepackage[caption=false]{subfig}
\usepackage{textcomp}
\usepackage{xcolor}
\usepackage{times}
\usepackage{soul}
\usepackage{url}
\usepackage[hidelinks]{hyperref}
\usepackage[utf8]{inputenc}
\usepackage{algorithm}
\usepackage{algorithmic}
\usepackage{multicol}
\usepackage{multirow}
\usepackage{rotating}

\usepackage{xspace}

\title{\LARGE \bf
SaViD: Spectravista Aesthetic Vision Integration for Robust and Discerning 3D Object Detection 
in Challenging Environments
}

\author{Tanmoy Dam$^{1}$,  Sanjay Bhargav Dharavath$^{2}$, Sameer Alam$^{1}$, Nimrod Lilith$^{1}$,  Aniruddha Maiti $^{3}$\\ Supriyo Chakraborty$^{2}$  and Mir Feroskhan$^{1}$ 
\thanks{*The authors would like to thank the funding support by the Start-Up Grant from the School of Mechanical and Aerospace Engineering at Nanyang Technological University.}
\thanks{${1}$ Tanmoy Dam, Sameer Alam, Nimrod Lilith, and Mir Feroskhan are associated with the Saab-NTU Joint Lab,
        Nanyang Technological University, Singapore
        {\tt\small tanmoydam@yahoo.com, (sameeralam, nimrod.lilith, mir.feroskhan)@ntu.edu.sg}. Tanmoy Dam and Mir Feroskhan are corresponding authors.}
\thanks{${2}$ Sanjay Bhargav Dharavath and Supriyo Chakraborty are associated with the Indian Institute of Technology, 
        Kharagpur, India
        {\tt\small (sanjay810, supriyochakraborty)@iitkgp.ac.in}}
\thanks{${3}$ Aniruddha Maiti, Assistant Professor, West Virginia State University {\tt\small aniruddha.maiti87@gmail.com}}
}

\begin{document}

\maketitle
\thispagestyle{empty}
\pagestyle{empty}

\begin{abstract}

The fusion of LiDAR and camera sensors has demonstrated significant effectiveness in achieving accurate detection for short-range tasks in autonomous driving. However, this fusion approach could face challenges when dealing with long-range detection scenarios due to disparity between sparsity of LiDAR and high-resolution camera data. Moreover, sensor corruption introduces complexities that affect the ability to maintain robustness, despite the growing adoption of sensor fusion in this domain. We present SaViD, a novel framework comprised of a three-stage fusion alignment mechanism designed to address long-range detection challenges in the presence of natural corruption. The SaViD framework consists of three key elements: the Global Memory Attention Network (GMAN), which enhances the extraction of image features through offering a deeper understanding of global patterns; the Attentional Sparse Memory Network (ASMN), which enhances the integration of LiDAR and image features; and the KNNnectivity Graph Fusion (KGF), which enables the entire fusion of spatial information. SaViD achieves superior performance on the long-range detection Argoverse-2 (AV2) dataset with a performance improvement of 9.87\% in AP value and an improvement of 2.39\% in mAPH for L2 difficulties on the Waymo Open dataset (WOD). Comprehensive experiments are carried out to showcase its robustness against 14 natural sensor corruptions. SaViD exhibits a robust performance improvement of 31.43\% for AV2 and 16.13\% for WOD in RCE value compared to other existing fusion-based methods while considering all the corruptions for both datasets. Our code is available at \href{https://github.com/sanjay-810/SAVID}{\textcolor{blue}{SaViD}}.

\end{abstract}

\begin{IEEEkeywords}
GMAN, ASMN, KGF, Multi-modal fusion, 3D object detection
\end{IEEEkeywords}

\section{INTRODUCTION}

Autonomous driving uses LiDAR and cameras for 3D object detection \cite{arnold2019survey,xiao2023collaborative,kamath2024physics,xiao2024real, chen2024design,dam2022developing}. Cameras offer high-resolution details; LiDAR adds depth and shape data. The fusion of both is crucial for accuracy but challenging\cite{arnold2019survey, 10610908}. This study addresses these fusion challenges with a robust solution.
\begin{figure}[h]
    \centering
    \begin{minipage}{0.20\textwidth}
        \includegraphics[width=\textwidth, height=2.2cm]{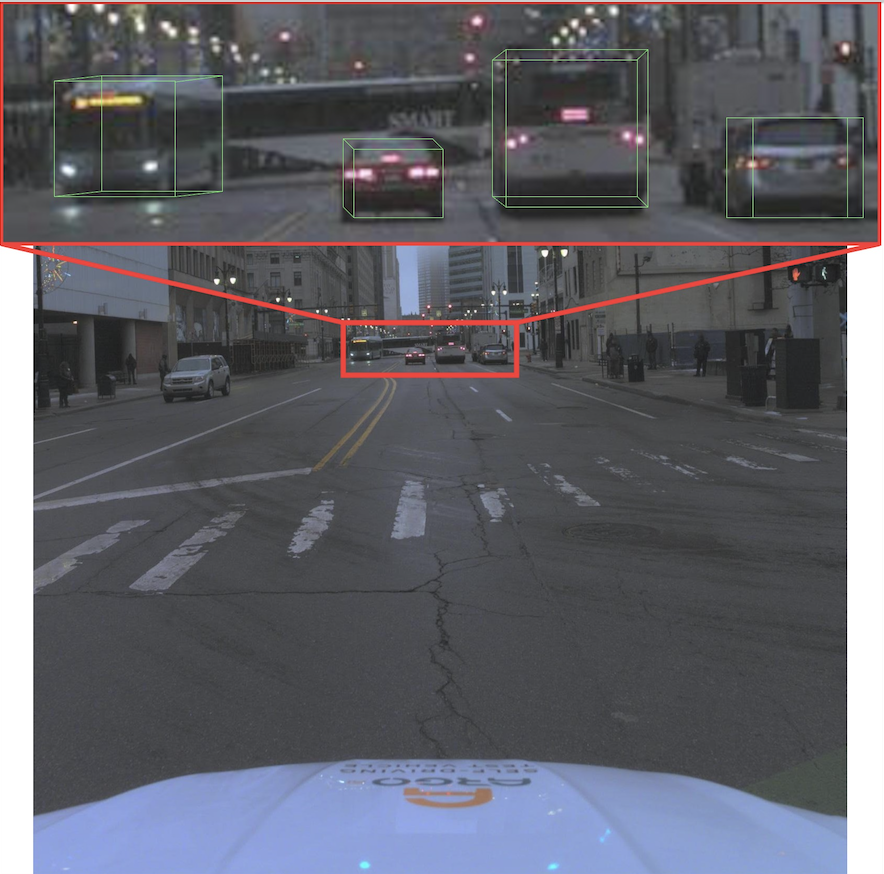}
        \caption*{a. Long-range objects}
    \end{minipage}
    \hfill
    \begin{minipage}{0.18\textwidth}
        \includegraphics[width=\textwidth, height=2.2cm]{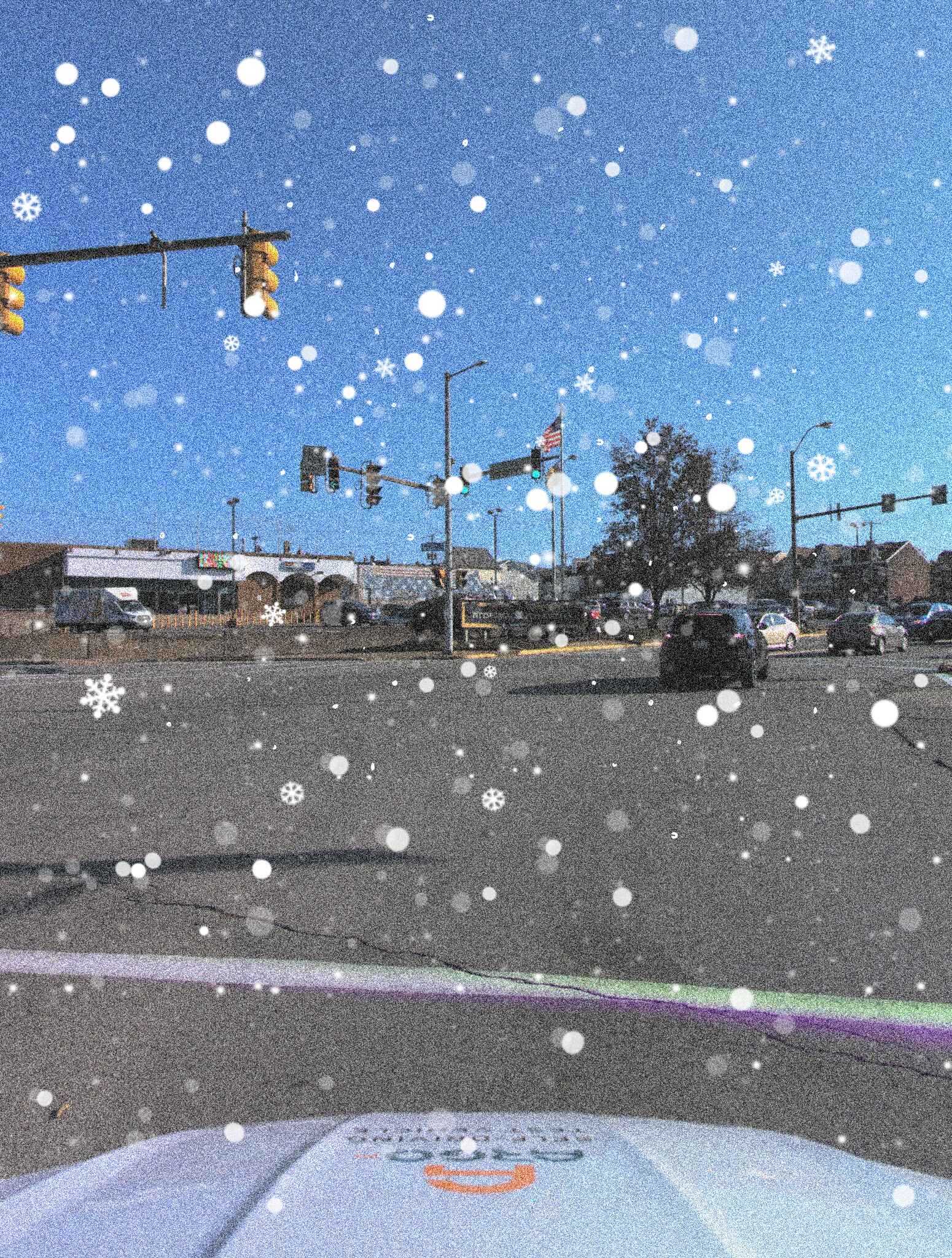}
        \caption*{b. Corruption-induced}
    \end{minipage}
    \caption{Argoverse-2 dataset two challenges}
    \label{fig:av2-challenges}
\end{figure}

\textbf{\textit{Challenge 1:}} \textit{Weak fusion or alignment between LiDAR and camera features for long-range detection.} While camera features are often fused with LiDAR data \cite{vora2020pointpainting}, most methods \cite{li2022deepfusion, bai2022transfusion, liang2022bevfusion} use mid-level fusion, such as additive fusion in AVOD \cite{ku2018joint} or cross-attention in DeepFusion \cite{li2022deepfusion} and TransFusion \cite{bai2022transfusion}. These approaches may miss long-range details due to low-resolution camera inputs and sparse LiDAR data, highlighting the need for strong fusion alignment for accurate 3D detection in long-range scenarios \cite{yin2021center}.

\textbf{\textit{Challenge 2:}} \textit{Robustness in long-range detection.} 

Data-driven deep learning models struggle to generalize on corrupted data from adverse weather, sensor noise, and other factors \cite{kilic2021lidar, hendrycks2019benchmarking, dong2023benchmarking}. This limits the reliability of autonomous driving. Recent robustness assessments have developed datasets focused on adverse conditions \cite{dong2023benchmarking}, but evaluations are mainly on small-range datasets like KITTI and nuScenes. Achieving robustness for long-range detection remains a significant challenge, requiring dedicated benchmark analysis.

\textbf{Our contribution.} We introduce SaViD, a novel method for robust long-range detection, distinct from traditional LiDAR-camera fusion techniques. SaViD achieves strong modality alignment by integrating sparse LiDAR point cloud data with camera features through local-global view representations \cite{yang2018ipod}, and incorporates natural robustness to effectively handle long-range detection and adverse conditions (Figure \ref{fig:av2-challenges}). To summarize, our main
contributions in this paper are described as follows:
\begin{itemize}
\item GMAN: A memory-based vision transformer that extracts image features using depth as a global query.
\item ASMN: A single-stage method for aligning sparse point cloud features with global image features.
\item KGF: A parameter-free fusion alignment technique for accurate integration of pseudo-point clouds and images.
\item SaViD achieves state-of-the-art performance on long-range detection in both clean and corrupted Argoverse-2 and Waymo Open Dataset.
\end{itemize}
\begin{figure*}
    \centering
    \scalebox{0.50}{
    \includegraphics{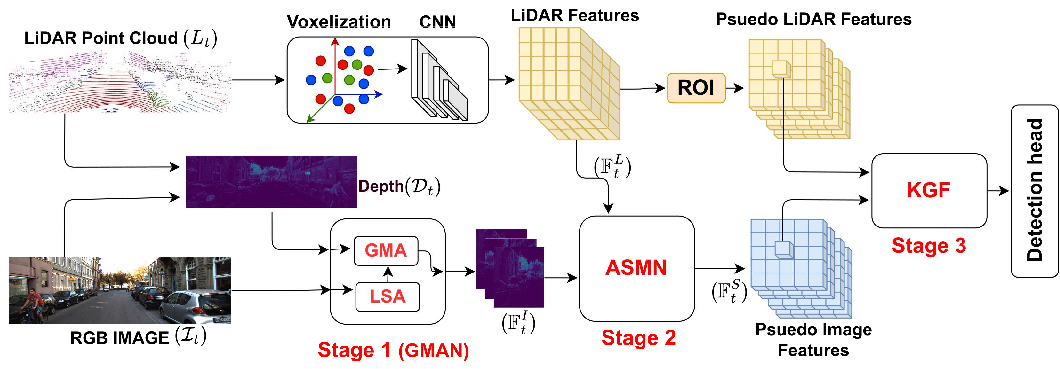}}
    \caption{SaViD Pipeline: The pipeline integrates multiple modalities through three essential components: GMAN, ASMN, and KGF. GMAN considers the image($\mathcal{I}_t$) as a local feature and the Depth($\mathcal{D}_t$) information as a global query, enabling an understanding of the scene through its vision-based Transformer in Stage 1. In Stage 2, a single-stage fusion integration of the ASMN module aligns the extracted image feature ($\mathbb{F}^{\mathcal{I}}_t$) by GMAN with the voxelized LiDAR feature, resulting in a cohesive global feature alignment. Finally, a parameter-free KGF alignment LiDAR RoI feature ($\mathbb{F}^{L}_t$) and the extracted ASMN Image feature ($\mathbb{F}^{S}_t$) with local importance are in Stage 3.}
    \label{fig:rce}
\end{figure*}

\section{Related Works}
\textbf{LiDAR Point Clouds for 3D Object Detection.} LiDAR-only 3D detection aims to predict 3D bounding boxes within raw, unprocessed point clouds. Detectors often project these points onto grids like pillars \cite{lang2019pointpillars}, range images \cite{yin2021center}, or 3D voxels \cite{zhou2018voxelnet} to handle their irregular structure. Neural networks such as PointNet \cite{qi2017pointnet} and PointNet++ \cite{qi2017pointnet++} extract features in the BEV plane, preserving object dimensions. Other methods use high-density range images for depth data \cite{yin2021center, bewley2020range, bhargav2024quantum}. These approaches can be classified as fully-dense, semi-dense, or point-dense \cite{fan2022fully, chen2017multiview}. Due to the sparse nature of point clouds, achieving strong performance in long-range detection with single-modality detectors remains challenging.

\textbf{LiDAR-camera Integration for 3D Object Detection.} LiDAR-camera fusion is challenging due to their differing data: LiDAR provides 3D depth, while cameras capture 2D visuals. Effective integration requires robust algorithms. Prior works like DeepFusion \cite{li2022deepfusion}, TransFusion \cite{bai2022transfusion}, and BEVFusion \cite{liang2022bevfusion} address these complexities. DeepFusion uses self-attention but struggles with long-range detection due to resolution mismatches between LiDAR and camera data. TransFusion’s Multi-head Attention faces issues with data density variations. BEVFusion’s simple concatenation of both modalities may hinder detection of distant objects due to blurred camera images and sparse LiDAR points. These limitations suggest that current attention-based techniques may not fully ensure effective fusion.

\section{SaViD Pipeline}

\textbf{Problem Definition.} The objective of this paper is to develop a robust 3D object detection approach using multi-modal sensors, achieving effective performance in challenging conditions. We consider multi-modal input-output sequences defined as $(\mathcal{X}_t, \mathcal{Y}_t) = \{(\mathcal{I}_t, L_t), (\mathcal{I}_{(t-1)}, L_{(t-1)}), \ldots\}$, where $\mathcal{I}_t \in \mathbb{R}^{H \times W \times 3}$ is the $t$-th camera image and $L_t \in \mathbb{R}^{N \times 3}$ is the LiDAR point cloud. The output $\mathcal{Y}_t$ consists of 3D bounding boxes associated with classes $\mathcal{M} = \{1, 2, \ldots, M\}$. Our framework, denoted as $\Phi$, predicts $\hat{\mathcal{Y}}_t = \Phi(\mathcal{I}_t, L_t)$ with high-confidence 3D bounding boxes, closely resembling the ground truth $\mathcal{Y}_t$. We assume the use of depth information $\mathcal{D}_t$, which can be estimated from $\mathcal{I}_t$ and $L_t$, as an additional input for $\Phi$.

\textbf{Depth Estimation ($\mathcal{D}_t$).} To generate a high-resolution depth map from sparse LiDAR data $L_t$, combined with correlated RGB imagery $\mathcal{I}_t$, we extract global features beneficial for image processing. Therefore, a frame of point clouds $L_t$ can be transformed into a sparse depth map $\mathcal{D}_t \in \mathbb{R}^{H \times W \times 3}$ using a projection function $\mathcal{T}_{L_t, \mathcal{I}_t} \rightarrow \mathcal{D}_t$. In this context, the mapping function $\mathcal{T}$, implemented as a depth neural network, uses both the RGB image $\mathcal{I}_t$ and point clouds $L_t$ to produce the high-resolution depth map $\mathcal{D}_t$.

\subsection{LiDAR Feature Extraction through Voxelization}
The point clouds $L_t$ are sparse and unevenly distributed. We preprocess by voxelizing the $t$-th point cloud with dimensions $H_v \times W_v \times C_v$ and compute voxel features by averaging point-wise features in non-empty voxels \cite{shi2023pv}. Key-points are identified using Furthest Point Sampling (FPS) \cite{shi2023pv}, selecting $\mathcal{K} = 4096$ key-points ($L_t^{\mathcal{K}}$) for experiments. The characteristics of non-empty voxels are obtained by averaging 3D coordinates and reflectance values of contained points. Feature volumes are transformed through $3 \times 3 \times 3$ 3D sparse convolutions, resulting in downsampled resolutions of $1\times, 2\times, 4\times$, and $8\times$. These volumes are represented as feature vectors assigned to individual voxels, with the final voxel feature vectors denoted as $\mathbb{F}_t^L \in \mathbb{R}^{H \times W \times C}$.

\subsection{Multi-Modal Feature Fusion Alignment} 
\begin{figure}[h]
    \centering
    \includegraphics[width=0.4\textwidth, height=4.5cm]{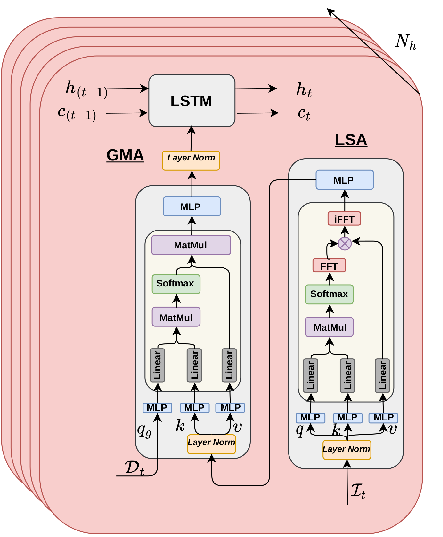}
    \caption{The GMAN Architecture is composed of both LSA and GMA blocks, employing a total of $N_{h}$ attention heads. Moreover, the LSTM block operates across consecutive time frames.}
    \label{fig:suuply_GMAN}
\end{figure}

\textbf{Stage 1: GMAN through $\mathcal{D}_t$.} We introduce GMAN, which combines local-global attention with frequency domain information via FFT-iFFT layers. As shown in Figure \ref{fig:suuply_GMAN}, the Local-Spectral  Attention is a vision-based transfer architecture \cite{hatamizadeh2022global}. The local image tensor $\mathcal{I}_t$ serves as the query with dimensions ${(B, H/P \times W/P, P \times P, C)}$, where $B$ is the batch size, $P \times P$ represents the local patch window size, and $C$ is the channel count. The aggregated batch size is $B^* = H/P \times W/P$. This paper proposes a novel transformer block that extracts features from $\mathcal{I}_t$ and $\mathcal{D}_t$ using two key modules: Local Spectral Attention (LSA) and Global Memory Attention (GMA). LSA and GMA capture local and global depth-relevant features to handle varying object scales within the same window. The local feature extraction is akin to the Swin Transformer \cite{liu2022swin} and Global Vision Transformer \cite{hatamizadeh2022global}. $\mathcal{I}_t$ passes through a local query generator utilizing the LSA module for detail-focused feature extraction. 

\textbf{GMA:} While LSA focuses on local patches within the current frame ($\mathcal{I}_t$), GMA operates on a unified framework, utilizing depth modality from depth maps ($\mathcal{D}_t$). Unlike single modality methods \cite{liu2023bevfusion}, GMA's global query computation is predefined, using globally extracted query tokens from $\mathcal{D}_t$ and interacting with local key and value representations from LSA. This allows GMA to integrate both local and global information. The GMA module leverages global context by applying attention across $\mathcal{I}_t$ and $\mathcal{D}_t$, correlating $\mathcal{I}_t$-derived key-value pairs with $\mathcal{D}_t$ as the query. Since $\mathcal{D}_t$ integrates LiDAR ($L_t$) and RGB ($\mathcal{I}_t$) data, GMA effectively attends to various locations within $L_t$, enhancing contextual understanding. GMA is formulated as:
\begin{equation}\label{eq:GMA}
\text{GMA}(\mathcal{I}_t, \mathcal{D}_t) = \mathbb{LN}(\alpha v), \quad \alpha = \text{Softmax}(g(q_g, k))
\end{equation}

where $q_g \in \mathcal{D}_t$, $k, v \in \mathcal{I}_t$ are the query, key, and value, respectively, and $g(\cdot)$ is the global attention function. Temporal feature accumulation is managed using ReLU-activated LSTM cells \cite{hochreiter1997long} after GMA to capture sequential patterns. Algorithm \ref{alg:GMA} provides pseudo-code for the GMA module.
\begin{algorithm}
\caption{Global Memory Attention (GMA)}
\label{alg:GMA}
\begin{algorithmic}
\STATE \textbf{Input/Output:} (B*, N, C) where B* = B $\times$ N*, $N_h$: Attention Heads
\STATE \textbf{Initialization:}
\STATE \quad $k_m, q_m, v_m = nn.\text{Linear}(C, C)$
\STATE \quad \text{softmax} = $nn.\text{Softmax}(dim=-1)$
\STATE \quad \text{LSTM} = $nn.\text{LSTM}(.)$

\STATE \textbf{Forward}($\mathcal{I}_t, \mathcal{D}_t$):
\STATE \quad $k = k_m(\mathcal{I}_t)$, $q\_g = q_m(\mathcal{D}_t)$, $v = v_m(\mathcal{I}_t)$
\STATE \quad $k, q\_g, v = k.\text{view}(B*, N, N_h, -1).\text{permute}(0, 2, 1, 3)$
\STATE \quad $q\_k = \text{matmul}(q\_g, k.\text{transpose}(-2, -1))$
\STATE \quad $\text{attn} = \text{softmax}(q\_k)$
\STATE \quad $\text{attn\_v} = \text{matmul}(\text{attn}, v.\text{transpose}(-2, -1))$
\STATE \quad \textbf{return} $\text{LSTM}(\text{attn\_v}).\text{reshape}(B^*, N, C)$
\end{algorithmic}
\end{algorithm}

\textbf{Stage 2: ASMN.} ASMN introduces a novel temporal fusion mechanism between feature extractors $\mathbb{F}_t^{\mathcal{I}} \in \mathbb{R}^{H \times W \times C}$ and $\mathbb{F}_t^{L} \in \mathbb{R}^{H \times W \times C}$. Unlike GMAN, ASMN uses a single-stage integration that handles sequential information from both modalities. To address the sparsity of LiDAR voxel features ($\mathbb{F}_t^{L}$), we incorporate sparse attention \cite{zhang2021sparse} combined with LSTM cells to enhance fusion with $\mathbb{F}_t^{\mathcal{I}}$. Following GMA principles, voxel features act as global queries, interacting with key-value pairs from image features through sparse attention. This interaction generates a correspondence map linking $\mathbb{F}_t^{L}$ to regions in $\mathbb{F}_t^{\mathcal{I}}$, with LSTM states adapting to both modalities for effective sequence integration.
\begin{figure}[h]
    \centering
    \includegraphics[width=0.4\textwidth, height=4.5cm]{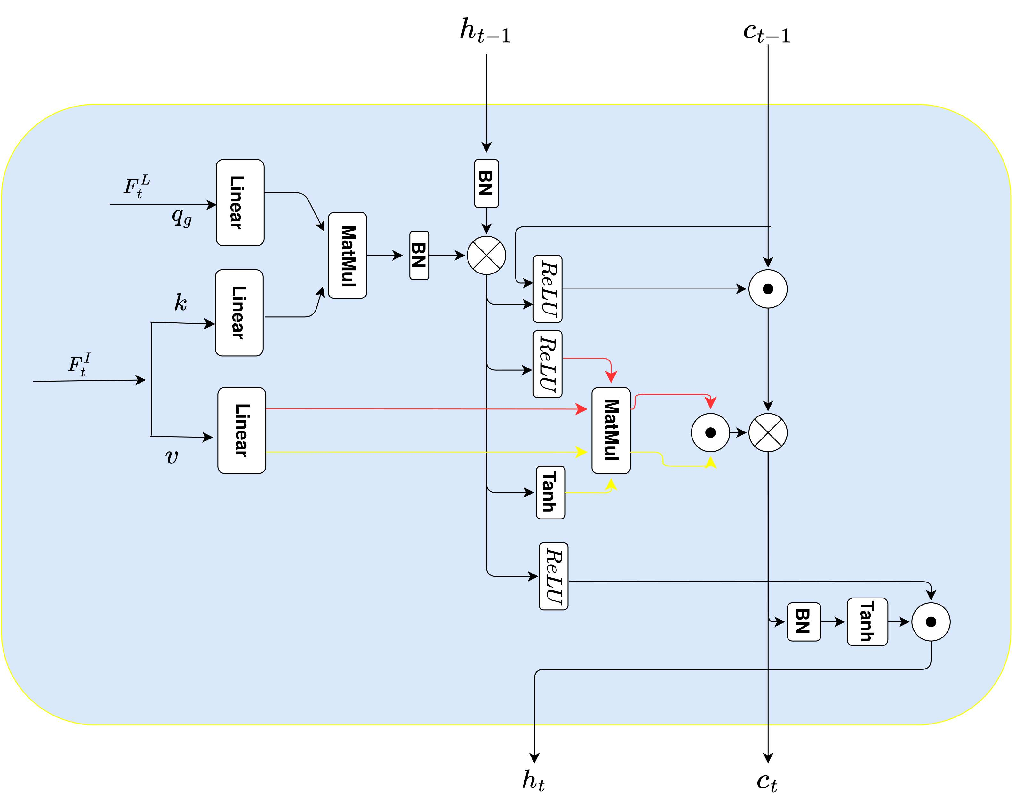}
    \caption{ASMN Architecture: The LiDAR feature extractor ($\mathbb{F}_t^{L}$) serves as a global query in a unified stage with $\mathbb{F}_t^{\mathcal{I}}$. Sequential data is processed using an LSTM cell with sparse attention and ReLU activation.} 
    \label{fig:suuply_ASMN}
\end{figure}

\begin{equation}\label{eq:GMA_final}
\text{ASMN}(\mathbb{F}_t^{\mathcal{I}}, \mathbb{F}_t^{L}) = (\beta_c v) \cdot \text{Tanh}(\beta_h v)
\end{equation}
where $\beta_h = f_h(q_g, k, h_{t-1})$ and $\beta_c = f_c(\beta_h, c_{t-1})$, with $q_g \in \mathbb{F}_t^{L}$, $k, v \in \mathbb{F}_t^{\mathcal{I}}$. $f_h$ and $f_c$ update key-query pairs using previous hidden and cell states with batch normalization and ReLU activation. Therefore, LSTM updates are:
\begin{equation}
c_t = \beta_c \cdot ((\beta_c v) \cdot \text{Tanh}(\beta_h v)), \quad h_t = \text{ReLU}(\beta_h) \cdot \text{Tanh}(c_t)
\end{equation}

For ASMN, $\beta_h$ and $\beta_c$ are defined as:
\begin{equation}\label{eq:ASMN_beta_h}
\beta_h = BN(\mathcal{l}(q_g) \cdot \mathcal{l}(k)) \cdot BN(h_{t-1}), \quad \beta_c = \text{ReLU}(\beta_h \cdot c_{t-1})
\end{equation}
\begin{equation}\label{eq:LSTM_GMAN}
\mathbb{F}_t^S = \text{ReLU}(\beta_c \cdot \mathcal{l} v) \cdot \text{Tanh}(\beta_h \cdot \mathcal{l} v)
\end{equation}

Here, $\mathcal{l}$ is a linear function, and $h_{t-1}$, $c_{t-1}$ are hidden and cell states from the previous time frame.\\
\textbf{Stage 3: KGF.} KGF is a parameter-free fusion alignment technique that captures local correlations between $\mathbb{F}_t^{S}$ and $\mathbb{F}_t^{L}$ by identifying similar attributes across modalities and emphasizing channel importance. Given ASMN-extracted features $\mathbb{F}_t^{S} \in \mathbb{R}^{H\times W\times C}$ and voxelized LiDAR features $\mathbb{F}_t^{L} \in \mathbb{R}^{H\times W\times C}$, KGF correlates pixel features $(\tau, \epsilon)$ with LiDAR points $(\xi, \gamma)$ using cosine distance, factoring in neighboring points. The minimum cosine distance $\mathbb{V}$ between corresponding features is calculated as:

\begin{equation}
    \text{Cosine}(\mathbb{F}_t^{S}, \mathbb{F}_t^{L})(\tau, \epsilon) = \min \left[ \frac{\mathbb{F}_t^{S} \cdot \mathbb{F}_t^{L}}{\sqrt{(\mathbb{F}_t^{S})^2 + (\mathbb{F}_t^{L})^2}} \right].
\end{equation}

The final KGF value accumulates weighted sums across channels and pseudo code is given 

\begin{equation}
    KGF(\tau, \epsilon, \kappa) = \sum_{\kappa=1}^{C} 2^{-\kappa} \cdot \mathbb{V}(\tau, \epsilon, \kappa) + \mathbb{F}_t^{S}(\tau, \epsilon, \kappa).
\end{equation}

Algorithm \ref{alg:KGF} for the pseudo-code of the KGF module.

\begin{algorithm}
\caption{KGF}
\label{alg:KGF}
\begin{algorithmic}
\STATE \textbf{Input:} $\mathbb{F}_t^L$, $\mathbb{F}_t^S$ : (H, W, C)
\STATE \textbf{Output:} Projected features

\STATE \textbf{def} \textbf{cosine\_dist}($a, b$):
\STATE \quad \textbf{return} $\frac{a \cdot b}{\sqrt{a^2 + b^2}}$

\STATE \textbf{def} \textbf{project}($\mathbb{F}_t^L, \mathbb{F}_t^S, \tau, \epsilon$):
\STATE \quad $ H\_range, W\_range, C = \text{shape}(\mathbb{F}_t^L)$
\STATE \quad $count = 0$
\STATE \quad \textbf{for} $\kappa$ \textbf{in} $C$:
\STATE \quad \quad $\text{val\_nei} = [\text{val} \mid \text{val} \in \text{neighbors}]$
\STATE \quad \quad $\text{min\_dist} = \min(\textbf{cosine\_dist}(\mathbb{F}_t^L, \mathbb{F}_t^S[\tau][\epsilon]) \mid (\tau, \epsilon) \in \text{val\_nei})$
\STATE \quad \quad $count += (2^{-\kappa}) \times \text{min\_dist}$
\STATE \quad \textbf{return} $count$

\STATE \textbf{def} \textbf{KGF}($\mathbb{F}_t^L, \mathbb{F}_t^S$):
\STATE \quad $output = \text{zeros\_like}(\mathbb{F}_t^S)$
\STATE \quad \textbf{for} $\kappa$ \textbf{in} $C$:
\STATE \quad \quad \textbf{for} $\tau$ \textbf{in} $H$, $\epsilon$ \textbf{in} $W$:
\STATE \quad \quad \quad $project\_value = \textbf{project}(\mathbb{F}_t^L, \mathbb{F}_t^S[\kappa], \tau, \epsilon)$
\STATE \quad \quad \quad $output[\kappa][\tau][\epsilon] = \mathbb{F}_t^S + project\_value$
\STATE \quad \textbf{return} $output$
\end{algorithmic}
\end{algorithm}

\subsection{Loss function} SaViD uses Voxel R-CNN \cite{deng2021voxel} for RPN and RoI loss, in addition to using Fusion Loss \cite{zhang2021sparse} and LSTM loss \cite{fan2023super}. 
\begin{table*}
\centering
\scalebox{0.70}{
\begin{tabular}{l|cccccccccccccccccccc|c}
\hline
Methods   &\begin{sideways}\textbf{Vehicle}\end{sideways} & \begin{sideways}\textbf{Bus} \end{sideways}  & \begin{sideways}\textbf{Pedestrian}\end{sideways} & \begin{sideways}\textbf{Stop Sign}\end{sideways} & \begin{sideways}Box Truck\end{sideways} & \begin{sideways}\textbf{Bollard}\end{sideways} & \begin{sideways}\textbf{C-Barrel}\end{sideways} & \begin{sideways}Motorcyclist\end{sideways} & \begin{sideways}\textbf{MPC-Sign}\end{sideways} & \begin{sideways}\textbf{Motorcycle} \end{sideways} & \begin{sideways}\textbf{Bicycle} \end{sideways} & \begin{sideways}\textbf{A-Bus} \end{sideways}& \begin{sideways}\textbf{School Bus} \end{sideways} & \begin{sideways}\textbf{Truck Cab} \end{sideways} & \begin{sideways}\textbf{C-Cone} \end{sideways}& \begin{sideways}\textbf{V-Trailer} \end{sideways} & \begin{sideways}\textbf{Sign} \end{sideways} & \begin{sideways}\textbf{Large Vehicle} \end{sideways} & \begin{sideways}\textbf{Stroller} \end{sideways} & \begin{sideways}\textbf{Bicyclist} \end{sideways} & \begin{sideways}\textbf{AP}\end{sideways}\\ 
\hline
\textit{Precision} &         &         &       &            &           &           &         &          &              &          &            &         &       &            &           &        &           &      &               &          &           \\ \hline
CenterPoint           & 61.0    & 36.0 & 33.0       & 28.0      & 26.0      & 25.0    & 22.5     & 16.0         & 16.0     & 12.5       & 9.5     & 8.5   & 7.5        & 8.0       & 8.0    & 7.0       & 6.5  & 3.0           & 2.0      & 14       & 17.5
    \\

CenterPoint\texttt{+}           & 67.6 & 38.9 & 46.5 & 16.9 & 37.4 & 40.1 & 32.2 & 28.6 & 27.4 & 33.4 & 24.5 & 8.7 & 25.8 & 22.6 & 29.5 & 22.4 & 6.3 & 3.9 & 0.5 & 20.1  & 26.67    \\

FSD                  & 67.1 & 39.8 & 57.4 & 21.3 & 38.3 & 38.3 & 38.1 & 30.0 & 23.6 & 38.1 & 25.5 & 15.6 & 30.0 & 20.1 & 38.9 & 23.9 & 7.9 & 5.1 & 5.7 & 27.0 & 29.58   \\ \hline


BEVFusion \texttt{\#}        & 67.2   &   39.8   &   58.1   &   31.9   &   36.3   &   35.2   &   36.7   &   34.1   &   26.1   &   46.8   &   33.6   &   21.2   &   22.2   &   16.9   &   31.2   &   22.8   &   13.2   &   5.4   &   9.6   &   32.6 & 31.05 \\ 

TransFusion \texttt{\#}          & 67.6 & 40.5 & 58.4 & 32.6 & 38.5 & 36.1 & 38.6 & 34.3 & 26.8 & 48.3 & 37.3 & 21.7 & 22.9 & 18.5 & 33.8 & 23.2 & 13.5 & 6.2 & 9.8 & 33.1  &  32.09  \\

DeepFusion \texttt{\#}         & 70.7 & 42.3 & 62.1 & 32.8 & 40.8 & 40.0 & 42.2 & 42.6 & 28.3 & 50.1 & 40.1 & 21.7 & 29.7 & 17.6 & 40.2 & 25.3 & 14.7 & 7.9 & 10.7 & 35.1 & 34.74     \\ \hline

SaViD (t=1)          & \textcolor{red}{78.2}                    & \textcolor{red}{48.6}                    & \textcolor{red}{67.8}                   & \textcolor{red}{38.6}                    & 40.7                     & \textcolor{red}{42.8}                     & \textcolor{red}{45.3}                     & 42.4                     & \textcolor{red}{30.8}                     & \textcolor{red}{53.2}                    & \textcolor{red}{41.5}                     & \textcolor{red}{25.9}                     & \textcolor{red}{30.9}                     & \textcolor{red}{22.6}                     & \textcolor{red}{41.3}                     & \textcolor{red}{30.9}                     & \textcolor{red}{19.6}                     & \textcolor{red}{10.8}                    & \textcolor{red}{12.8}                   &\textcolor{red}{38.8}    & \textcolor{red}{38.17 (+3.43)}  \\ 

SaViD (t=7)        & \textcolor{blue}{79.7}                     & \textcolor{blue}{49.5 }                    &\textcolor{blue}{ 68.7}                     & \textcolor{blue}{40.1 }                    & \textcolor{blue}{41.9}                     & \textcolor{blue}{43.8}                     & \textcolor{blue}{47.2 }                    & \textcolor{blue}{44.1}                     & \textcolor{blue}{33.3}                     & \textcolor{blue}{55.4 }                    & \textcolor{blue}{42.1}                     & \textcolor{blue}{25.9 }                    & \textcolor{blue}{32.3}                     & \textcolor{blue}{25.1}                     & \textcolor{blue}{44.9}                     & \textcolor{blue}{31.6 }                    & \textcolor{blue}{20.3 }                    & \textcolor{blue}{12.9 }                   & \textcolor{blue}{13.7 }                   &\textcolor{blue}{40.5}     &\textcolor{blue}{39.65(+1.48) } \\ \hline
\end{tabular}}
\caption{The table shows AP results on AV2-C validation for categories like C-Barrel, MPC-Sign, A-Bus, C-Cone, and V-Trailer. Bolded AP values indicate significant gains over single-frame ($t=1$) performance. \texttt{\#}: Simulated in the same environments.
}
\label{tab:AV2-C performance}
\end{table*}

\begin{figure*}
    \centering
    \scalebox{0.05}{
    \includegraphics{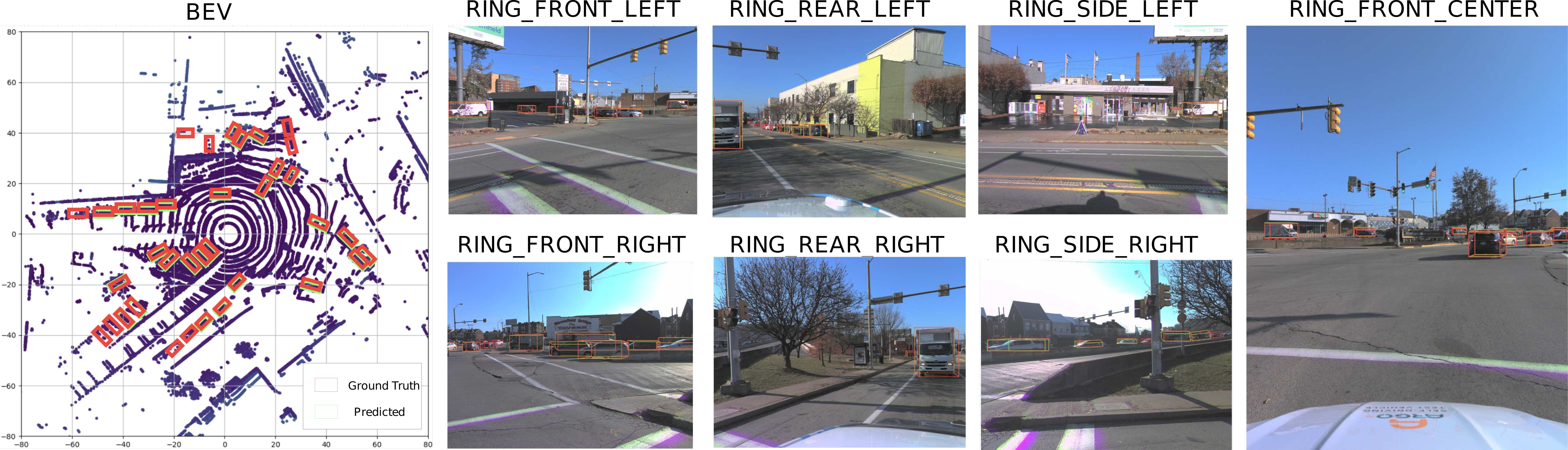}}
    \caption{A qualitative comparison of long-range 3D object detection using multi-modal fusion methods on AV2-C validation
set. BEV maps on left, 2D image space on right. Red: Ground truth, Green: Predicted boxes. }
    \label{fig:qualatative_performance_on AV2-C}
\end{figure*}

\begin{table*}[ht]
\centering
\caption{Comparison of Model Performance for 3D Detection on the WOD Test Set. In the table, `L' and `I' denote LiDAR and camera sensors, respectively. `TTA' and `Ens' represent test-time augmentation and ensemble model outputs, indicated by \texttt{\#}}
\label{tab:AYDIV_test_set}
\scalebox{1}{
\begin{tabular}{l|l|c|c|c|c}
\hline
\textbf{Method} & \textbf{Modality} & \textbf{ALL (mAPH)} & \textbf{VEH (APH)} & \textbf{PED (APH)} & \textbf{CYC (APH)} \\ \hline
SaViD (t=7)  (Ours) & L+I & \textcolor{blue}{82.96} \textcolor{red}{(+1.94)} & \textcolor{blue}{82.94} & \textcolor{blue}{84.15} & \textcolor{blue}{81.78} \\ 
SaViD (t=1) (Ours) & L+I & 82.16 & 82.37 & 83.51 & 80.59 \\ \hline
LoGoNet Ens\texttt{\#} \cite{li2023logonet} & L+I & 81.02 & 81.72 & 81.28 & 80.06 \\ 
BEVFusion TTA\texttt{\#} \cite{liu2023bevfusion} & L+I & 79.97 & 80.92 & 79.65 & 79.33 \\ 
LidarMultiNet TTA\texttt{\#} \cite{ye2023lidarmultinet} & L & 79.94 & 80.36 & 79.86 & 79.59 \\ 
MPPNet Ens\texttt{\#} \cite{chen2022mppnet} & L & 79.60 & 80.93 & 80.14 & 77.73 \\ 
MT-Net Ens\texttt{\#} \cite{chen2022mt} & L & 78.45 & 80.11 & 78.08 & 77.17 \\ 
DeepFusion Ens\texttt{\#} \cite{li2022deepfusion} & L+I & 78.41 & 79.09 & 78.57 & 77.58 \\ 
AFDetV2 Ens\texttt{\#} \cite{hu2022afdetv2} & L & 77.64 & 78.34 & 76.75 & 77.83 \\ 
INT Ens\texttt{\#} \cite{xu2022int} & L & 77.21 & 78.73 & 76.36 & 76.54 \\ 
HorizonLiDAR3D Ens\texttt{\#} \cite{ding20201st} & L+I & 77.11 & 77.83 & 76.50 & 76.98 \\ \hline
LoGoNet \cite{li2023logonet} & L+I & 77.10 & 79.30 & 78.91 & 73.10 \\ 
BEVFusion \cite{liu2023bevfusion} & L+I & 76.33 & 77.48 & 76.41 & 75.09 \\ 
CenterFormer \cite{zhou2022centerformer} & L & 76.29 & 78.28 & 77.42 & 73.17 \\ 
MPPNet \cite{chen2022mppnet} & L & 75.67 & 76.91 & 75.93 & 74.18 \\ 
DeepFusion \cite{li2022deepfusion} & L+I & 75.54 & 75.69 & 76.40 & 74.51 \\ \hline
\end{tabular}
}
\label{tab:Waymo_test_set}
\end{table*}

\section{Experiments}
\subsection{Dataset details}
Our goal is to conduct robust, long-range experiments with multi-modal fusion using the Argoverse2 (AV2) \cite{fan2022fully} and Waymo Open Dataset (WOD).

\textbf{AV2:} includes 1000 sequences: 700 for training, 150 for validation, and 150 for testing. It has a perception range of 200 meters and covers a 400m $\times$ 400m area, making it more extensive than other standard benchmarks like Waymo \cite{li2022deepfusion} and nuScenes \cite{bai2022transfusion}. AV2 features 30 object categories with a long-tail distribution; we focus on the top 20 classes, excluding the 10 tail classes. SaViD is tested in two scenarios: AV2-C for clean data and AV2-R (AV2-Robust) for corrupted sensor data ($\mathcal{L}_t, \mathcal{I}_t$).

\textbf{WOD} is the leading benchmark for LiDAR-based 3D object detection, known for its large and complex dataset of 1,150 sequences with over 200,000 frames, including LiDAR points, camera images, and 3D bounding boxes. The dataset is split into 798 training, 202 validation, and 150 testing sequences. The clean WOD-C detection range is 75 meters, covering a 150m x 150m area. Our evaluation focuses on long-range performance using \text{LEVEL\texttt{\string_}2 (L2)} difficulty, excluding \text{LEVEL\texttt{\string_}1 (L1)} for small-range detection. We also introduce WOD-R for corrupted sensors ($\mathcal{L}_t, \mathcal{I}_t$), similar to AV2-R.

\subsection{Natural Robustness} 
Natural robustness addresses real-world corruptions in autonomous driving, categorized into weather-induced and sensor-induced corruptions. We identify 14 common corruptions relevant to AV2 and WOD for long-range detection \cite{dong2023benchmarking}.

\textbf{Weather-Induced Corruptions.} These include Snow, Rain, Fog, and Sunlight, significantly affecting LiDAR and camera data. Weather effects are simulated on LiDAR using physics-based methods \cite{hahner2021fog,hahner2022lidar, kilic2021lidar} and visually augmented for cameras \cite{hendrycks2019benchmarking}.

\textbf{Sensor-Induced Corruptions.} We introduce 10 sensor-level corruptions: seven for LiDAR (e.g., Density Decrease, Cutout, LiDAR Crosstalk, FOV Lost, various noise types) and three for images (Gaussian, Uniform, and Impulse Noise) to simulate visual disturbances from lighting or camera faults \cite{hendrycks2019benchmarking}.

\subsection{Implementation Details}

\textbf{Network Architecture.} SaViD employs a three-stage strategy to integrate features from LiDAR and image modalities, using pseudo feature extraction. For LiDAR, SaViD builds on the Voxel-RCNN framework \cite{shi2020pv} with dynamic voxelization and feature dimensions of 16, 32, 64, and 64 to manage sparse point clouds. The image stream feature extractor relies on depth information ($\mathcal{D}_t$) from a pretrained Twise network \cite{imran2021depth}. In the GMA module, a dropout rate of 30\% is applied to the attention affinity matrix during training, with parameters: dimension = 64, $N_h$ = 8, and $P$ = 7. The MLP layer after GMA is a fully connected layer with 64 filters.

\textbf{Training and Inference Details.} SaViD is trained from scratch using the ADAM optimizer on 32 GTX 1080 Tesla T4 GPUs with a cosine annealing learning rate. The proposal refinement stage samples 128 proposals, maintaining a 1:1 ratio of positive (IoU $\geq$ 0.55) to negative proposals for enhanced long-range detection. Data augmentation techniques are employed during training \cite{shi2023pv, 10610908}. For inference, non-maximum suppression (NMS) is used in the RPN with IoU thresholds of 0.7 and 0.1 to filter redundant predictions \cite{shi2020pv}.

\subsection{Performance on AV2-C and WOD-C}
We evaluated single and multi-modal fusion methods on AV2-C (Table~\ref{tab:AV2-C performance}). Initial results using CenterPoint improved by 52.4\% in AP with the modified CenterPoint$\textit{+}$. The FSD method, with its sparse attention mechanism, enhanced performance by 10.91\% over CenterPoint$\textit{+}$. Two-modality 3D detection outperformed single-modality approaches, consistent with AV2-C results. BEVFusion improved AP by 4.73\% over FSD, and TransFusion surpassed BEVFusion by 3.34\%. DeepFusion, with cross-former feature alignment, further improved AP by 7.62\% over TransFusion. SaViD, using three-stage feature fusion, achieved the highest AP of 38.17, a 9.87\% gain over DeepFusion. Adding temporal alignment in SaViD increased AP to 39.65, up 3.73\% from the single-frame model.Figure~\ref{fig:qualatative_performance_on AV2-C} highlights SaViD’s qualitative performance using BEV maps and front camera views. 

In Table~\ref{tab:Waymo_test_set}, we compare model performance for 3D detection on the WOD test set for clean data. SaViD (t=7) achieved the highest mAPH of 82.96, improving L2 difficulties by 1.94 points. It excelled across all classes, with APH scores of 82.94 for vehicles, 84.15 for pedestrians, and 81.78 for cyclists. The single-frame SaViD (t=1) also outperformed LoGoNet Ens, with improvements of 1.14 mAPH, 1.38 APH for vehicles, 1.34 APH for pedestrians, and 1.56 APH for cyclists. Compared to BEVFusion \cite{liu2023bevfusion}, SaViD (t=7) showed a 2.99 mAPH increase, with specific gains of 2.02 for vehicles, 4.50 for pedestrians, and 2.45 for cyclists. The single-frame SaViD (t=1) also surpassed BEVFusion with a 2.19 mAPH boost. SaViD (t=7) outperformed LidarMultiNet TTA \cite{ye2023lidarmultinet} by 3.02 mAPH, and SaViD (t=1) achieved a 2.22 mAPH increase compared to LidarMultiNet TTA. These results underscore SaViD (t=7)'s superior performance in 3D object detection across various categories, establishing its effectiveness compared to other methods, especially when leveraging sequential frame information.

\subsection{Performance on AV2-R and WOD-R}
Robustness performance is assessed by measuring individual and relative corruption effects at various severity levels. Fusion-based methods' clean performance on AV2-C is denoted as $AP_{cln}$, while corrupted performances are denoted $AP_{r,s}$ for each corruption type (r) and severity level (s) \cite{dong2023benchmarking}. The average corruption performance is given by:

\begin{equation}
    AP_{corr} = \frac{1}{|\nu|} \sum_{r \in \nu} \frac{1}{5} \sum_{s=1}^{5} AP_{r,s},
\end{equation}

where $\nu$ represents the set of corruptions. The Relative Corruption Error (RCE) quantifies robustness by evaluating performance degradation under clean conditions:

\begin{equation}
    RCE = \left[  \frac{ AP_{cln} - AP_{corr} } { AP_{cln} }  \right], \quad RCE_{r,s} = \left[  \frac{ AP_{cln} - AP_{r,s} } { AP_{cln} }  \right].
\end{equation}

\begin{figure}
    \centering
    \subfloat[AV2-R]{
       \includegraphics[width=0.40\textwidth, height=2.5cm]{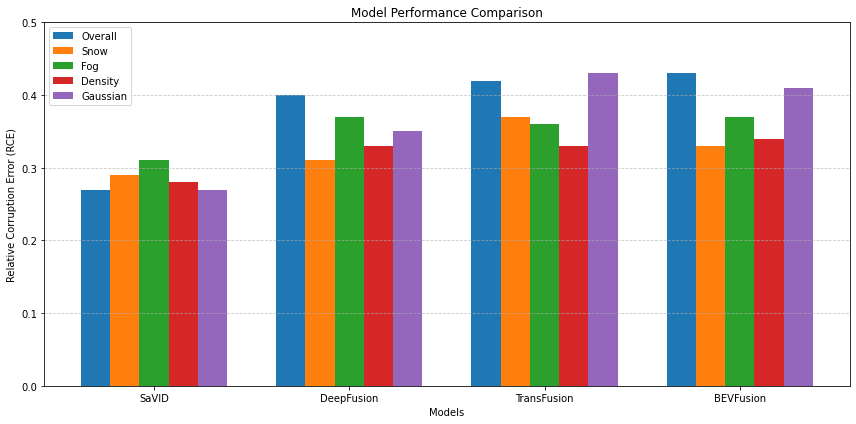}
        \label{fig:rce}
    } \\
    \subfloat[WOD-R]{
       \includegraphics[width=0.40\textwidth, height=2.5cm]{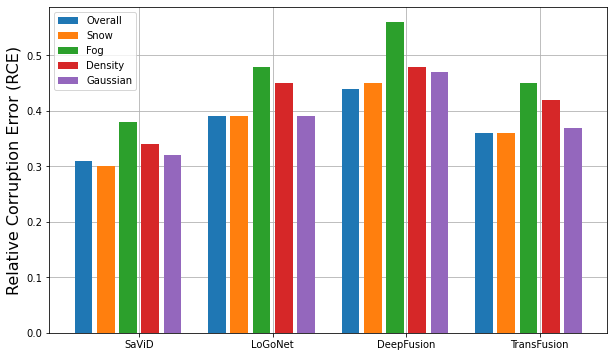}
        \label{fig:wod_rce}
    }
    \caption{The RCE depicts the overall results under all levels of corruption as well as the outcomes under each level of corruption for AV2-R and WOD-R dataset}
    \label{fig:robustness_RCE}
\end{figure}

Figure~\ref{fig:robustness_RCE} shows individual and overall RCE for AV2-R and WOD-R datasets. SaViD demonstrates the most robust performance with a 27.24\% overall RCE on AV2-R, a 31.43\% improvement over DeepFusion. BEVFusion shows the lowest robustness with a 48.82\% RCE decline. On WOD-R, SaViD's RCE drop is 31.24\%, compared to 36.36\% for LoGoNet, and 44.15\% for Transfusion. SaViD's superior robustness is due to its 3-stage fusion alignment mechanism.

\section{Ablation Study on AV2-C}
\textbf{Necessities of three-stage Fusions.} We conducted extensive experiments, as shown in Table~\ref{tab:ablation_studies}, to assess the impact of the three components on SaViD's performance: (1) Missing ASMN: while considering the only local-global image feature extractor GMAN and KGF, there is a 7.2\% decrease in AP value compared to considering all the components. The observed performance drop can be attributed to the limitations of the sparse LiDAR feature extractor($\mathbb{F}_t^{L}$). Despite estimating depth ($\mathcal{D}_{t}$) using LiDAR on Image, it fails to ensure optimal fusion alignment, leading to sub-optimal results in the fusion process. (2) Missing GMAN: while excluding the image feature extractor results in a 35\% decrease in the AP value compared to using all the components. Despite providing projected image data($\mathcal{I}_t)$ to align with the channels of the LiDAR feature extractor($\mathbb{F}_t^{L}$), it practically lacks information due to the sparsity of the LiDAR data. Consequently, it behaves like a conventional voxelized LiDAR-based detector. 
\begin{table}
\centering
\begin{tabular}{l|l|l|l}
GMAN & ASMN & KGF & AP \\ \hline
\checkmark & & \checkmark & 35.4 \\
& \checkmark & \checkmark & 24.8 \\
\checkmark & \checkmark & \checkmark & 38.17 \\ \hline
\end{tabular}
\caption{Performance of Each Component in SaViD (t=1)}
\label{tab:ablation_studies}
\end{table}

\textbf{Necessities of $\mathcal{D}_t$ in SaViD.} In our experiment, we examine two variations of vision transformers, namely SwinV2 \cite{liu2022swin} and GCVIT \cite{hatamizadeh2022global}. We integrate them with the two proposed fusion stages, ASMN and KGF, while excluding the depth information ($\mathcal{D}_t$). Table~\ref{tab: Depth Estimation performance on varying Vision transformer} presents the results. When using SwinV2 to extract $\mathbb{F}_t^{I}$, the obtained AP value is 33.4. However, when considering fused conv2D in GCVIT, the performance improved to 34.9 in AP. Nevertheless, both cases exhibit sub-optimal performance due to the lack of alignment between spatial feature extraction and sparse LiDAR information. Despite the inclusion of other two proposed alignment methods, the absence of global context LiDAR information (disparity with image) leads to a notable decline in long-range detection performance. Therefore, this ablation study highlights the importance of using $\mathcal{D}_t$ as a global query to minimize the disparity with $\mathcal{I}_t$ and improve performance.

\begin{table}[h]
\centering
\begin{tabular}{l|c}
\hline
Model & AP  \\ \hline
SwinV2\textbackslash \(\mathcal{D}_t\) + ASMN + KGF & 33.4 \\
GCVIT\textbackslash \(\mathcal{D}_t\) + ASMN + KGF & 34.9 \\
GMAN + ASMN + KGF (t=1) & 38.17 \\ \hline
\end{tabular}
\caption{Performance Without $\mathcal{D}_t$ on Varying Vision Transformers}
\label{tab: Depth Estimation performance on varying Vision transformer}
\end{table}

\section{CONCLUSION}
This paper introduces SaViD, a novel three-stage robust fusion alignment method incorporating a local-global perspective for 3D object detection. The first stage uses a vision transformer-based GMAN to extract image features, considering local and global depth information. It then introduces a ASMN to align sparse LiDAR features with extracted image features. Lastly, a parameter-free KGF achieves final fusion. SaViD achieves notable performance gains on AV2-C, shows resilience to corruptions on AV2-R, and excels on WOD, especially in L2 difficulties. With its long-range detection capability, SaViD's potential extends beyond AVs to Digital Airport Tower Control, enhancing operational efficacy and safety in complex airport environments.

\bibliographystyle{IEEEtran}
\bibliography{icdm}

\end{document}